\documentclass[floatfix,pra,amsmath,amssymb,letterpaper,groupaddresses,superscriptaddress,twocolumn]{revtex4-1}
\usepackage{times}
\usepackage{latexsym}
\usepackage{graphicx}
\usepackage{amsmath,amssymb}
\usepackage{verbatim,times,bbm}
\usepackage{subfigure}
\usepackage{soul}
\usepackage{float}
\usepackage[colorlinks,hyperindex]{hyperref}
\hypersetup{colorlinks,citecolor=blue,linkcolor=blue,urlcolor=blue}

\usepackage{color}

\newcommand{\gre}{\color{green}}

\begin{document}
	
\title{Mpemba Effect in Parametrically Driven Coupled Oscillators under White and Colored Noise}
	
\author{Aref Pahlevani\footnote{a.pahlevani@shahroodut.ac.ir}}
\affiliation{Faculty of Physics, Shahrood University of Technology, 3619995161 Shahrood, Iran}
	
\author{Morteza Rafiee\footnote{m.rafiee@shahroodut.ac.ir}}
\affiliation{Faculty of Physics, Shahrood University of Technology, 3619995161 Shahrood, Iran}
	
\author{Mehdi Ansari-rad\footnote{ansari.rad@shahroodut.ac.ir}}
\affiliation{Faculty of Physics, Shahrood University of Technology, 3619995161 Shahrood, Iran}
	
\begin{abstract}
We study the Mpemba effect in a pair of linearly coupled harmonic oscillators, one of which is parametrically driven and coupled to an independent thermal bath. Using the covariance-matrix formalism, we derive the relaxation dynamics under both Gaussian white noise and Lorentzian colored noise, including single-channel and dual-channel noise embeddings. We characterize relaxation through the Frobenius distance to the steady state and through the projection onto the slowest mode of the dynamical generator. Our results show that parametric driving provides the primary control knob for anomalous relaxation: as the drive approaches the boundary of the stable region explored in this work, the Mpemba crossing time decreases systematically. Colored noise further enhances the effect, with dual-oscillator Lorentzian noise producing a stronger reduction in the crossing time than single-oscillator noise and enlarging the parameter region where the Mpemba effect occurs. Nevertheless, the slow-mode structure of the drift matrix remains the dominant mechanism, while the influence of colored noise is secondary and mainly quantitative. We show that the Mpemba crossing time decreases as the system approaches the parametric stability boundary and that Lorentzian colored noise enlarges the region in the parametric and coupling strength plane where the Mpemba effect is observed.
	
\end{abstract}

\maketitle
	
\section{Introduction}
	
Understanding relaxation toward thermal equilibrium is one of the central topics in non-equilibrium statistical physics. 
Typically, it is expected that the physical systems initially closer to equilibrium relax faster than those starting farther away. It has been shown that some hotter systems can reach equilibrium faster than colder ones under special conditions. This effect was originally reported by Erasto Mpemba in the cooling of water and later analyzed in non-equilibrium statistical physics. Generally, the situation in which a system initially at higher temperature relaxes faster than one prepared at a lower temperature is called the Mpemba effect. This phenomenon has been investigated in various classical systems\cite{Jeng2006,Lasanta2017,Lapolla220,Kumar20202,Bechhoefer2021,Biswas2023}. Quantum Mpemba effects have also been studied in quantum systems, with attention to both open and isolated dynamics\cite{Filiberto Ares}. 
	
In recent years, many studies have been conducted to understand the origins of the Mpemba effect. In the systems whose dynamics can be decomposed into multiple relaxation modes, anomalous relaxation dynamics can arise and the relaxation dynamics can be expressed in terms of eigenstates of the evolution operator. The Mpemba effect can occur whenever the initial hotter state has smaller overlap with the slowest relaxation mode. \cite{Lu2017,Hirschberg2019}.
This effect has also been demonstrated in a variety of complex systems, including spin models, glass dynamics, and open quantum systems \cite{Baity2019,Strachan2025}. 	
Several studies on the quantum Mpemba effect have shown that it can be engineered by exploiting the sub-block structure of the dynamical generator in thermalizing open Markovian systems \cite{Mattia Moroder}. Moreover, the quantum Mpemba effect has been demonstrated in a quantum dot coupled to two reservoirs described by the Anderson model\cite{Amit Kumar Chatterjee}.

For more general systems, the Pontus–Mpemba effect has been introduced as a two-step protocol that includes the preparation time of a system in a distant initial state for an arbitrary non-equilibrium configuration \cite{Andrea Nava}. Quantum Pontus–Mpemba effects have also been studied under both real-time and imaginary-time dynamics in systems with U(1) symmetry and ferromagnetic initial states \cite{Hui Yu}.
In addition, the Mpemba effect has been investigated in the context of quantum oscillators and two-level systems
\cite{Kheirandish} and the reader is referred to a comprehensive review of this effect for both experimental and theoretical studies in classical and quantum systems \cite{Teza2026}.
	
Many of the open quantum systems that have been studied involve white noise and Markovian dynamics, whereas the role of environmental memory and finite correlation times has received less attention. One way to address these effects is to consider noise processes with Lorentzian spectral density, which arise naturally in stochastic descriptions of non-equilibrium systems \cite{gardiner2004,Kampen1992}.
Understanding how environmental memory influences relaxation dynamics is therefore important for assessing the Mpemba effect in realistic systems.
	
Open quantum harmonic oscillators with parametric amplification provide a particularly interesting platform for studying non-equilibrium relaxation phenomena. Such systems can be regarded as bosonic modes that evolve into squeezed-thermal states and appear in several physical contexts, including cavity optomechanics, excited resonators, and quantum amplification devices \cite{Huang2009, Shahidani2022,Bothner2020,Burd2019,Lemonde2016,Marti2024,Royer2018,Wu2024}. The effective oscillator frequencies and relaxation spectrum are significantly altered by the parametric amplification process and consequently, such systems provide an externally controlled relaxation pathway. 	
In this work, we investigate the Mpemba effect in a parametrically driven dissipative oscillator system consisting of two linearly coupled harmonic oscillators dissipated into distinct thermal baths. Using a covariance matrix approach, we derive the governing relaxation equations under Gaussian white noise and Lorentzian colored noise applied to one or both momentum quadrature channels, and we analyze the spectral structure of the corresponding evolution operator. 
Our results reveal three main findings: (i) the parametric driving amplitude $\Lambda$ provides effective external control over the Mpemba crossing time $t_W^*$ under white Gaussian noise, with $t_L^*$ under single colored Lorentzian noise or $t_{LL}^*$ under double colored Lorentzian noise decreasing systematically as the system approaches the stability boundary of the parameter regime considered in this work; (ii) Lorentzian colored noise applied simultaneously to both oscillators reduces the crossing time more efficiently than single-oscillator noise and extends the parameter regime where the Mpemba effect occurs to values of $\Lambda$ farther from the boundary of the parameter domain explored in this work; and (iii) although colored noise enhances the effect, the dominant mechanism remains the deterministic spectral structure of the drift matrix, as evidenced by the robustness of the slow-mode projections. Although the crossing time depends only moderately on the squeezing strength $r$ and the squeezing angle $\phi$ once a crossing exists, a minimum squeezing strength is required to prepare the initial states that exhibit the Mpemba crossing. These findings demonstrate that parametric amplification offers a powerful knob for engineering anomalous relaxation in experimentally accessible platforms such as cavity optomechanics and superconducting circuits, where non-Markovian noise provides a measurable but secondary enhancement to the primary spectral control mechanism.
While the Mpemba effect has been investigated in a variety of classical and quantum systems, including open quantum oscillators and non-Markovian environments, the combined influence of parametric driving and colored environmental fluctuations on Mpemba relaxation has not been explored. The present work introduces a unified covariance-matrix framework that treats both Gaussian white-noise and Lorentzian colored-noise environments within the same spectral-relaxation formalism. This approach allows us to identify the respective roles of deterministic drift dynamics and environmental memory in anomalous relaxation. In particular, we demonstrate that parametric driving serves as the primary control parameter governing the Mpemba crossing time, whereas colored noise mainly enhances the effect by enlarging the parameter region in which Mpemba relaxation occurs. These results provide a systematic analysis of how externally controlled driving and non-Markovian fluctuations jointly influence relaxation pathways in coupled bosonic systems.
Parametric driving of bosonic modes is routinely implemented in cavity optomechanical systems and superconducting parametric circuits \cite{Aspelmeyer2014,Castellanos2007}. Recent optomechanics experiments with narrowband noisy drives have revealed clear bandwidth-dependent deviations from a Markovian description, making them relevant examples of colored-noise dynamics \cite{Banniard2024}. Meanwhile, controlled non-Markovian dynamics has already been demonstrated experimentally in open quantum systems \cite{Liu2011}. These developments suggest that the interplay between parametric driving and environmental memory predicted in the present work should be experimentally accessible.
	
The remainder of the paper is organized as follows. In Sec. \ref{sec:model}, we introduce the parametrically driven oscillator model and derive the corresponding Langevin equations. In Sec. \ref{sec: spectral}, we present the covariance matrix formalism and analyze the spectral structure of the relaxation dynamics. Section \ref{sec:numeric} discusses the numerical results demonstrating the Mpemba effect and investigating  the dependence of the crossing time and Mpemba strength on the parametric driving strength and noise correlation time. Finally, Sec. \ref{sec:conclusion} summarizes our conclusions.
	
	
\section{The model}\label{sec:model}
We consider a system consisting of two linearly coupled harmonic modes, each dissipating into an independent thermal bath, one of which is parametrically driven. In the rotating frame, the total Hamiltonian can be written as ($\hbar=1$),
	\begin{align}
	H_0 &=\omega_a \hat{a}^\dagger \hat{a} + \Delta_b \hat{b}^\dagger \hat{b} + \Lambda (\hat{b}^2  + \hat{b}^{\dagger 2}  )\\ \nonumber
	&+ g (\hat{a} + \hat{a}^\dagger)(\hat{b} + \hat{b}^\dagger)  
\end{align}

where $a (a^\dagger)$ and $b (b^\dagger)$ are the annihilation (creation) operators of the harmonic oscillators and $\omega_a$ is the frequency of mode $a$, while $\Delta_b$ is the effective detuning of mode $b$ in the rotating frame. Parametric amplification of the harmonic oscillator with strength $\Lambda$ has been shown as the third term in the Hamiltonian and the strength of the coupling of these two oscillators is denoted by $g$.
Dynamics of an open quantum system can be obtained by master equations or quantum Langevin equations \cite{Breuer2002, Gardiner2004}. In Gaussian bosonic systems, such as optical cavities, optomechanical systems, and parametric amplifiers, dynamics can be fully described using covariance matrices \cite{Weedbrook2012,Serafini2017}.
For the quadrature vector
\begin{equation}
	\mathbf{R}=(x_a,p_a,x_b,p_b)^T,
\end{equation}
where, $x_a=(\hat{a} + \hat{a}^\dagger)/\sqrt{2}$, $p_a=(\hat{a} - \hat{a}^\dagger)/i \sqrt{2}$ and $x_b$ and $p_b$ are defined similarly by substituting $\hat{a}$ with $\hat{b}$.  
The Langevin equations with white noise are
\begin{align}
	\dot x_a &= -\gamma_a x_a + \omega_a p_a + \sqrt{2\gamma_a}\,x_{a,{\rm in}}(t), \\
	\dot p_a &= -\omega_a x_a - \gamma_a p_a + Gx_b + \sqrt{2\gamma_a}\,p_{a,{\rm in}}(t), \\
	\dot x_b &= -\gamma_b x_b + \delta p_b + \sqrt{2\gamma_b}\,x_{b,{\rm in}}(t), \\
	\dot p_b &= Gx_a - \Delta x_b - \gamma_b p_b + \sqrt{2\gamma_b}\,p_{b,{\rm in}}(t).
\end{align}
where, $G=2g$, $\delta=\Delta_b - 2 \Lambda$, $\Delta=\Delta_b + 2 \Lambda$ and $\gamma_a (\gamma_b)$ is the damping rate of oscillator $a (b)$. The corresponding white-noise correlations are
\begin{align}
	\langle x_{a,{\rm in}}(t)x_{a,{\rm in}}(t')\rangle
	&=
	\langle p_{a,{\rm in}}(t)p_{a,{\rm in}}(t')\rangle
	=
	\frac{1}{2}(2\bar n_a+1)\delta(t-t'), \\
	\langle x_{b,{\rm in}}(t)x_{b,{\rm in}}(t')\rangle
	&=
	\langle p_{b,{\rm in}}(t)p_{b,{\rm in}}(t')\rangle
	=
	\frac{1}{2}(2\bar n_b+1)\delta(t-t').
\end{align}
The dynamics obeys linear quantum Langevin equations \begin{equation}
	\dot{\mathbf{R}}(t) = A\mathbf{R}(t) + \boldsymbol{\xi}(t),
\end{equation}
where $\boldsymbol{\xi}(t)$ is the vector of Gaussian noise and the drift matrix is
\begin{equation}
	A =
	\begin{pmatrix}
		-\gamma_a & \omega_a & 0 & 0 \\
		-\omega_a & -\gamma_a & G & 0 \\
		0 & 0 & -\gamma_b & \delta \\
		G & 0 & -\Delta & -\gamma_b
    \end{pmatrix}.
\end{equation}
The covariance matrix is defined as
\begin{equation}
	\sigma_{ij}
	=\frac{1}{2}
	\langle R_iR_j + R_jR_i \rangle
	-\langle R_i\rangle\langle R_j\rangle .
\end{equation}
Its time evolution follows the Lyapunov equation
\begin{equation}
	\dot{\sigma}
	=A\sigma + \sigma A^T + D .
\end{equation}
The diffusion matrix $D=diag\{\gamma_a(2 \bar n_a+1), \gamma_a(2 \bar n_a+1), \gamma_b(2\bar n_b+1), \gamma_b(2\bar n_b+1)\}$ describes environmental noise where  occupation numbers in terms of covariance matrix elements define as $n_j(t)=\frac{1}{2}\left[	\sigma_{x_jx_j}(t) +	\sigma_{p_jp_j}(t) - 1	\right]$, $\qquad	j=a,b$.
The steady state covariance matrix is obtained from
\begin{equation}\label{steady cov}
	A\sigma_{ss} + \sigma_{ss}A^T + D = 0 .
\end{equation}
The Routh–Hurwitz criterion for $(\omega_a, \delta)>0 $ which implies $\Lambda < 0.5$, gives the stability condition $(\omega_a^2 + \gamma_a^2)(\gamma_b^2 + \delta \Delta) - G^2 \delta \omega_a >0 $ \cite{Routh_Hurwitz, Shahidani2022},  see also Appendix \ref{App:Routh_Hurwitz}. 
To model realistic dephasing environments, we next apply colored noise to the $p_a$ quadrature where momentum fluctuations exhibit memory. In this case, the symmetry between position and momentum of subsystem $a$ is broken, while remaining dynamics are still under Markovian embedding. This kind of colored noise can be applied via structured photonic or phononic baths in optomechanical setups.
For Ornstein--Uhlenbeck colored noise acting only on the $p_a$ channel, the white-noise term in the $p_a$ equation is replaced by a colored force $\eta(t)$ \cite{Hanggi1995, Lien2025}:
\begin{align}
	\dot x_a &= -\gamma_a x_a + \omega_a p_a + \sqrt{2\gamma_a}\,x_{a,{\rm in}}(t), \\
	\dot p_a &= -\omega_a x_a - \gamma_a p_a + Gx_b  + g_L \eta(t), \\
	\dot x_b &= -\gamma_b x_b + \delta p_b + \sqrt{2\gamma_b}\,x_{b,{\rm in}}(t), \\
	\dot p_b &= Gx_a - \Delta x_b - \gamma_b p_b + \sqrt{2\gamma_b}\,p_{b,{\rm in}}(t).
\end{align}
with
\begin{equation}
	\dot{\eta}(t)=-\gamma_L \eta(t)+\sqrt{2D_0}\,\xi(t),
\end{equation}
where $\xi(t)$ is a standard white noise source, which produces Lorentzian noise with spectrum
\begin{equation}
	S(\omega) =
	\frac{2 D_0}{\omega^2 + \gamma_L^2}
\end{equation}
The corresponding correlation function is 
\begin{equation}
	\langle \eta(t)\eta(t')\rangle = \frac{D_0}{\gamma_L}e^{-\gamma_L|t-t'|}.
\end{equation}
The effective force is defined as
\begin{equation}
	\xi_L(t)=g_L\,\eta(t),
\end{equation}
with
\begin{equation}
	\langle \xi_L(t)\xi_L(t')\rangle
	= g_L^2 \frac{D_0}{\gamma_L}e^{-\gamma_L|t-t'|}.
\end{equation}
A convenient Markovian embedding is obtained by introducing the extended vector
\begin{equation}
	\mathbf{R}_{\rm ext}=(x_a,p_a,x_b,p_b,\eta)^T,
\end{equation}
with drift matrix
\begin{equation}
	A_{\rm ext}=
	\begin{pmatrix}
		-\gamma_a & \omega_a & 0 & 0 & 0 \\
		-\omega_a & -\gamma_a & G & 0 & g_L \\
		0 & 0 & -\gamma_b & \delta & 0 \\
		G & 0 & -\Delta & -\gamma_b & 0 \\
		0 & 0 & 0 & 0 & -\gamma_L
	\end{pmatrix},
\end{equation}
And the diffusion matrix is $D=diag\{\gamma_a(2 \bar n_a+1), \gamma_a(2\bar n_a+1), \gamma_b(2\bar n_b+1), \gamma_b(2\bar n_b+1), 2D_0\}$.
	
Applying the same construction to both oscillators yields
\begin{equation}
	\mathbf{R}_{\rm ext}=(x_a,p_a,x_b,p_b,\eta_a,\eta_b)^T,
\end{equation}
with drift matrix
\begin{equation}
	A_{\rm ext}=
	\begin{pmatrix}
		-\gamma_a & \omega_a & 0 & 0 & 0 & 0 \\
		-\omega_a & -\gamma_a & G & 0 & g_{La} & 0\\
		0 & 0 & -\gamma_b & \delta & 0 & 0\\
		G & 0 & -\Delta & -\gamma_b & 0 & g_{Lb}\\
		0 & 0 & 0 & 0 & -\gamma_{La} & 0\\
		0 & 0 & 0 & 0 & 0 & -\gamma_{Lb}
	\end{pmatrix},
\end{equation}
And the diffusion matrix is $D=diag\{\gamma_a(2\bar n_a+1), \gamma_a(2 \bar n_a+1), \gamma_b(2\bar n_b+1), \gamma_b(2\bar n_b+1), 2D_{0a}, 2D_{0b}\}$.
	
\section{Spectral Relaxation Theory of the Mpemba Effect} \label{sec: spectral}
In non-equilibrium statistical physics, relaxation toward equilibrium or steady state is one of the central concepts of interest. A useful way to analyze relaxation dynamics is through the spectral decomposition of the dynamical generator governing the evolution of the system \cite{Breuer2002,Prosen2010}.
For our model whose dynamical equation is the covariance matrix evolution, the deviation from the steady state is defined by
\begin{equation}
	\delta\sigma(t) = \sigma(t) - \sigma_{ss},
\end{equation}
and its dynamics becomes
\begin{equation}
	\dot{\delta\sigma}	=	A\delta\sigma + \delta\sigma A^T .
\end{equation}
So, Vectorizing the covariance matrix leads to find the covariance generator as
\begin{equation}
	\mathcal{L} = A\otimes I + I\otimes A .
\end{equation}
where, $\mathcal{L}$ is the dynamical generator and its eigenvalues determine the relaxation spectrum of the covariance matrix as
\begin{equation}
	\mathbf{\delta\sigma}(t) =	\sum_k c_k e^{\lambda_k t} \mathbf{v}_k,
\end{equation}
where the relaxation modes are determined by eigenvalues $\lambda_k$ and eigenvectors $\mathbf{v}_k$ and the projections of the initial state onto the left eigenvectors $\mathbf{w}_k$ of $\mathcal{L}$ are given by the coefficients $c_k = \langle \mathbf{w}_k , \mathbf{x}(0) \rangle$. The slowest relaxation mode of the system is defined by the eigenvalue with the largest real part, i.e., the one closest to zero.
\begin{equation}
	\lambda_1 = \max_k \mathrm{Re}(\lambda_k)
\end{equation}
At long times the dynamics is dominated by this mode,
\begin{equation}
	\mathbf{\delta\sigma}(t) \approx c_1 e^{\lambda_1 t} \mathbf{v}_1 .
\end{equation}
This phenomenon is known as spectral relaxation and plays a fundamental role in non-equilibrium dynamics \cite{Kampen1992}.
The Mpemba effect can also be understood within this framework \cite{Lu2017,Klich2019, Kessler2022}. Suppose two initial hot and cold states are expanded as
\begin{equation}
	\delta\sigma_h(0) = \sum_k c_k^{(h)} v_k , \hspace{5mm} 	\delta\sigma_c(0) = \sum_k c_k^{(c)} v_k.
\end{equation}
At long times, relaxation is again dominated by the slowest eigenvalue,
\begin{equation}
	\delta\sigma(t) \approx c_1 e^{\lambda_1 t} v_1 .
\end{equation}
A Mpemba effect occurs when
\begin{equation}
	|c_1^{(h)}| < |c_1^{(c)}|.
\end{equation}
In other words, if the initially hotter state has a smaller projection onto the slowest relaxation mode, it can reach equilibrium faster than the colder state. Thus, the relaxation rates of the two states are determined by their overlaps with the slow relaxation manifold in state space.
To quantify relaxation, we use the Frobenius distance, which is a natural and widely used matrix norm for comparing covariance matrices. Frobenius-norm-based measures have also been used to characterize state coherence and asymmetry \cite{Yao2016} as
\begin{equation}
	\mathcal{D}(\sigma(t)) = \sqrt{\mathrm{Tr}[(\sigma(t)-\sigma_{\mathrm{ss}})^T (\sigma(t)-\sigma_{\mathrm{ss}})]}.
\end{equation}
In our analysis, we identify the Mpemba effect using two complementary criteria: (i) the Frobenius distance cross time, $\mathcal{D}(\sigma_\mathrm{h}(t^*)) = \mathcal{D}(\sigma_\mathrm{c}(t^*))$ with $\mathcal{D}(\sigma_\mathrm{h}(0)) > \mathcal{D}(\sigma_\mathrm{c}(0))$, and (ii) the projection amplitude onto the slowest mode $\lambda_1 = \max_k \mathrm{Re}(\lambda_k)$ which satisfies $|c_1^{(\mathrm{h})}| < |c_1^{(\mathrm{c})}|$. Both criteria are verified numerically in Sec. \ref{sec:numeric}.
	
\section{numerical results}\label{sec:numeric}
 we employed the exact matrix-exponential solution of the Lyapunov equation,
\begin{equation}
	\sigma(t) = e^{At} \left[\sigma(0)-\sigma_{\rm ss}
	\right]	e^{A^T t} +	\sigma_{\rm ss},
\end{equation}
.
We consider two different initial Gaussian states for the oscillators, $\sigma_{th}^{(i)}(0)= Blockdiag \{ \frac{1}{2}(2 \bar n_a^i+1)I, \frac{1}{2}(2 \bar n_b^i + 1)I \}$, where $i=c, h$ correspond to cold and hot states, respectively.  
To prepare the nonequilibrium initial conditions used in this work, we apply the same squeezing transformation to both the hot and cold thermal states  with identical squeezing parameters but different thermal occupations:

\begin{equation}
	\sigma_c(0)	= \left[S(r,\phi)\otimes I \right]
	\sigma_{\rm th}(\bar n_c)\left[	S(r,\phi)\otimes I
	\right]^T,
\end{equation}
\begin{equation}
	\sigma_h(0)	=\left[S(r,\phi)\otimes I \right]
	\sigma_{\rm th}(\bar n_h)\left[	S(r,\phi)\otimes I
	\right]^T,
\end{equation}
with $\bar n_h>\bar n_c$.

Thus, squeezing is applied equally to both initial states.
where $S(r,\phi)$ is the squeezing matrix
\begin{equation}
	S(r,\phi)=
	\begin{pmatrix}
	\cosh r -\cos\phi\sinh r & -\sin\phi\sinh r \\
	-\sin\phi\sinh r & \cosh r +\cos\phi\sinh r
	\end{pmatrix}.
\end{equation}
		
In this work, squeezing is used only as an initial-state preparation tool and does not modify the subsequent system dynamics.
Moreover, to compare two initial states within the same family, we choose two squeezed thermal states with the same squeezing parameters but different temperatures as the hot and cold initial states. This choice allows us to test whether nonclassical structure in the initial state can accelerate relaxation beyond that of a colder thermal reference state.
In all calculations, we use $\omega_a = \Delta_b =1$, $\bar n_a=1$, $\bar n_b=5$, $\gamma_a = \gamma_b = 0.45 \omega_a$, $G=0.2 \omega_a$, $\Lambda = 0.48 $, $r=1$ {\gre ,} $\phi=\pi/4$, $\bar{n}_a^c (0) = \bar{n}_b^c(0) = 2 $ and $\bar{n}_a^h (0) = \bar{n}_b^h(0) = 7 $. The colored noise with Lorentzian spectrum is characterized by $\gamma_L = 0.09 \omega_a$, $D_0 = 0.3 \omega_a$ and $ g_L = 0.32 \omega_a$. These values correspond to the weak noise regime with the dimensionless parameters as $g_L^2 D_0/(\gamma_L \omega_a) \leq min(\gamma_a,\gamma_b)$. In addition, the stability condition $ \mathrm{Re}(\lambda_i)<0 $, where $\lambda_i$ are the eigenvalues of the drift matrix $A$, has been checked in all numerical calculations.
	
\begin{figure}
	\begin{center}
		\includegraphics[width=8.5cm, height=7.5cm]{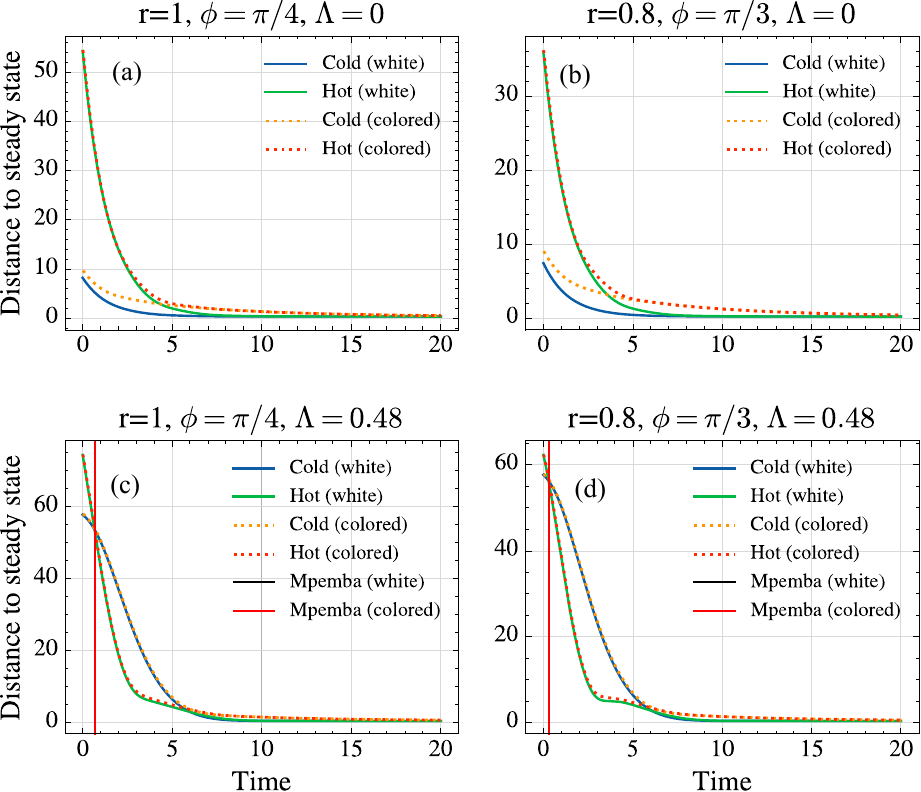}
    \end{center}
	\caption{ 
			(color online) Time evolution of the Frobenius distance ($D(\sigma(t)$) to the steady state for hot and cold initial states under Gaussian white noise, (a) and (b) without parametric driving ($\Lambda=0$) while squeezing parameters $r=1$, $\phi=\pi/4$ ($r=0.8$, $\phi=\pi/3$), respectively. The same results obtained with $\Lambda=0.48$ in (c) and (d).}\label{fig1}
\end{figure}

 As a first step, we examined the dependence of the Mpemba	crossing time on the squeezing parameters. The results,
presented in Appendix \ref{App:Squeezing parameters}, show that a crossing is observed only for sufficiently large squeezing strengths, approximately $r \ge 0.6$ for the parameters considered here.
Figure \ref{fig1}(a) and (b) show the time evolution of the Frobenius distance to the steady state under both white in the absence of parametric driving ($\Lambda=0$) and  squeezing parameters $r=1$, $\phi=\pi/4$ ($r=0.8$, $\phi=\pi/3$), respectively. In this case, the weak colored noise does not significantly affect the dynamics relative to white noise, and no Mpemba crossing is observed for both cases. The same evolution, shown in Fig.\ref{fig1}(c)-(d) with nonzero parametric drive, $\Lambda=0.48$, and different squeezing parameters $r=1$, $\phi=\pi/4$ ($r=0.8$, $\phi=\pi/3$), respectively, demonstrate the occurrence of the Mpemba effect.
Since, the qualitative behavior of the system is similar for
all suitable squeezing parameters satisfying $r \ge 0.6$,
we use $r=1$ and $\phi=\pi/4$ in the remainder of the paper.

\begin{figure}
	\begin{center}
		\subfigure{\label{Lambda0}
			\includegraphics[width=4.cm,height=3.5cm]{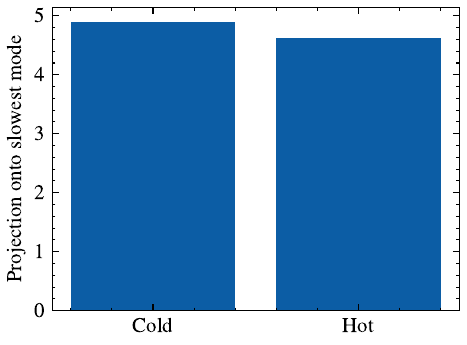}}\subfigure{\label{Lambda048}
			\includegraphics[width=4cm,height=3.5cm]{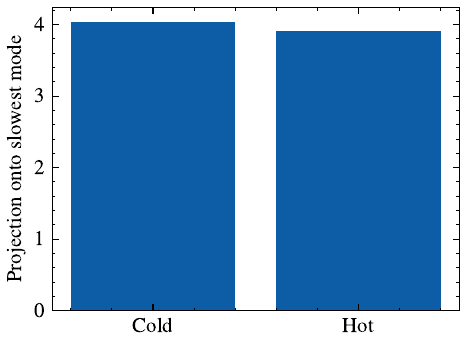}}
	\end{center}
	\caption{(color online) Projection amplitudes onto the slowest relaxation mode (\(|c_1|\)) for hot and cold initial states under white (left) and colored noise (right). Hotter states have smaller projection amplitudes (\(|c_1^{(h)}| < |c_1^{(c)}|\)), confirming the occurrence of the Mpemba effect.}\label{fig2}
\end{figure}

To verify the Mpemba effect more directly, we also plot the projections onto the slowest relaxation mode under white and weak colored noise in Fig. \ref{fig2} where parametric driving is turned on ($\Lambda=0.48$). The results yield $|c_1^{(c)}| = 4.896, |c_1^{(h)}| = 4.632$ and $|c_1^{(c)}| = 4.038, |c_1^{(h)}| = 3.912$ for white and colored noises respectively, confirming that the hotter state has the smaller overlap with the slowest mode in the presence of parametric drive.

\begin{figure}
	\begin{center}
		\includegraphics[width=8cm,height=6.25cm]{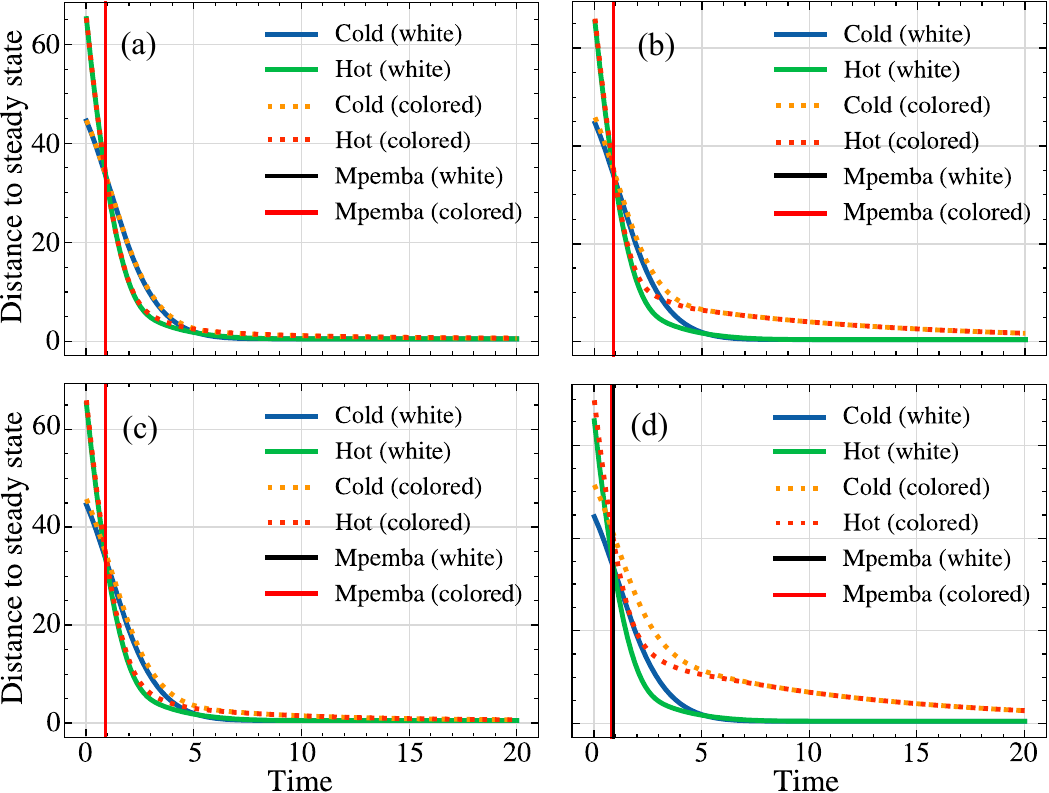}
	\end{center}
	\caption{(color online) Time evolution of the Frobenius distance $D(\sigma(t))$ to the steady state for hot and cold initial states under Gaussian white noise and Lorentzian colored noise at $\Lambda=0.48$. (a) and (b) compare white noise with single-oscillator colored noise in the weak ($\gamma_L=0.09 \omega_a$, $g_L=0.32 \omega_a$, $D_0=0.3 \omega_a$) and moderate noise regimes ($\gamma_L=0.05 \omega_a$, $g_L=0.5 \omega_a$, $D_0=0.4 \omega_a$), respectively. (c) and (d) compare white noise with dual-oscillator colored noise in the weak and moderate noise regimes, respectively. In all cases, colored noise shortens the Mpemba crossing time relative to white noise, with the dual-noise configuration producing the strongest acceleration of relaxation.}\label{fig3}
\end{figure}

To explore stronger memory effects, we next increase the colored noise parameters to $\gamma_L=0.05 \omega_a$, $D_0=0.4 \omega_a$ and $g_L=0.5 \omega_a$.
Figure (\ref{fig3}) compares the Frobenius distance dynamics for white and colored noise in the presence of parametric driving $\Lambda=0.48$. In the weak-noise regime, the single-channel Lorentzian noise produces only a modest reduction in the Mpemba crossing time relative to white noise, indicating that environmental memory has a limited but visible effect. When the colored noise strength is increased, the separation between the hot and cold trajectories becomes more pronounced and the crossing occurs earlier, showing that stronger non-Markovian correlations accelerate the relaxation process. This trend is even clearer when colored noise acts simultaneously on both oscillators. The dual-noise configuration reduces the crossing time more efficiently than the single-noise case and yields the fastest approach to the steady state among the cases considered. These results confirm that colored noise enhances anomalous relaxation, while the dominant control of the effect still comes from the parametric drive and the underlying spectral structure of the drift matrix.
	
Our numerical analysis shows that as the parametric drive approaches the boundary of the stable parameter region considered here, the Mpemba crossing time decreases.For the parameter regime considered here, this behavior is primarily associated with changes in the eigenvector structure, mode non-normality, and slow-mode projections rather than a substantial shift in the real part of the slowest eigenvalue. This behavior demonstrates that parametrically driven systems provide a platform for enhancing and controlling relaxation phenomena such as the Mpemba effect especially since such systems appear in several experimental platforms including cavity optomechanics, superconducting microwave circuits, and driven mechanical resonators.
	
\begin{figure}
	\begin{center}
			\includegraphics[width=8.5cm, height=7.25cm]{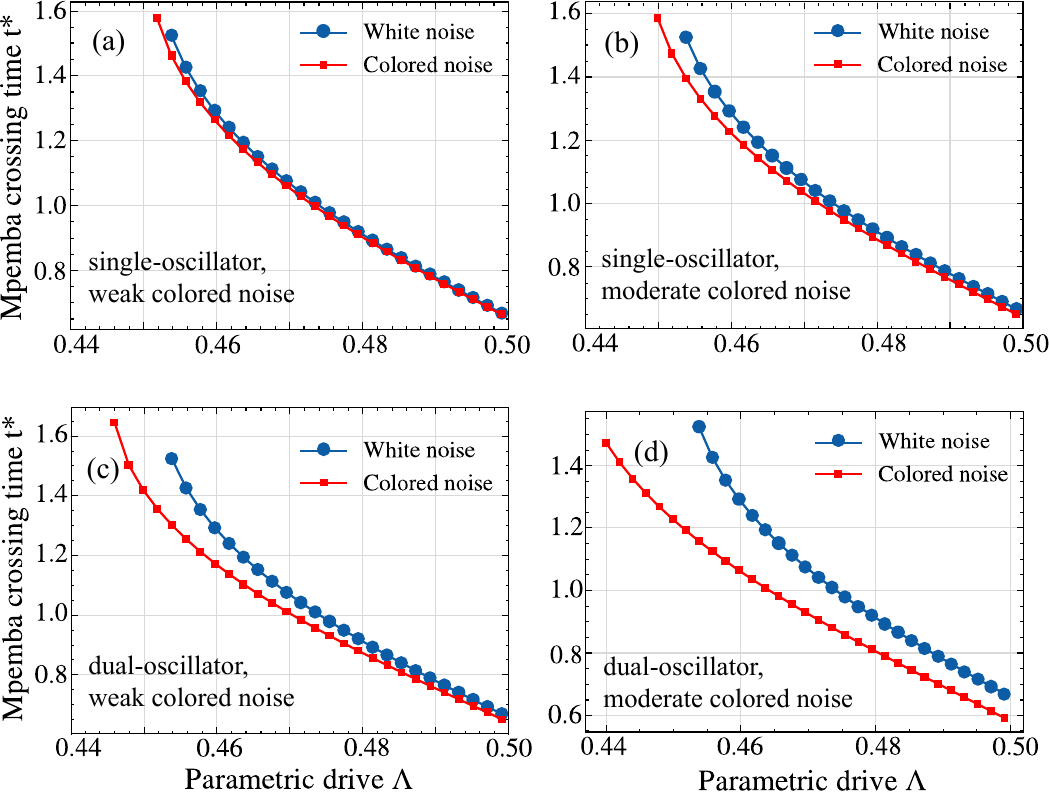}
	\end{center}
	\caption{(color online)  Mpemba crossing time $t^*$ as a function of the parametric drive $\Lambda$ under Gaussian white noise and Lorentzian colored noise. (a) and (b) show the single-oscillator weak and moderate colored noise cases, respectively, while (c) and (d) show the dual weak and moderate colored noise cases, respectively. Blue circles denote white-noise results and red squares denote colored noise results. Colored noise reduces the crossing time more efficiently than white noise and extends the Mpemba regime toward smaller values of $\Lambda$, away from the boundary of the stable parameter regime considered here, $\Lambda_c=0.5$.}\label{fig4}
\end{figure}

Figure~\ref{fig4} presents comprehensive phase diagrams of the Mpemba crossing time $t^*$ versus parametric driving $\Lambda$ in both weak ($\gamma_L=0.09\omega_a$, $g_L=0.32 \omega_a$ and $D_0=0.3 \omega_a$) and moderate ($\gamma_L=0.05\omega_a$, $g_L=0.5 \omega_a$ and $D_0=0.4 \omega_a$) noise regimes. Lorentzian colored noise applied simultaneously to both oscillators reduces $t^*$ more effectively than white noise over the entire stable parameter range as the parameter strength moves away from the boundary of the parameter regime considered in this work ($\Lambda_c=0.5$). Moreover, colored noise extends the Mpemba regime to significantly smaller values of $\Lambda$, where white noise fails to produce crossings. For example, at $\Lambda=0.44$, white noise does not yield any crossing time while dual colored noise yields $t^*=1.46$ (moderate noise). Also, comparison with single-oscillator colored noise confirms that distributed memory effects in both subsystems provide the strongest improvement. These results indicate that non-Markovian correlations not only accelerate anomalous relaxation, but also expand the parameter space available for Mpemba effect.
	
Meanwhile, Table \ref{tabletimes} summarizes all Mpemba crossing times $t^*$, where $t^*_{LL}$ denotes the dual-oscillator colored noise case. The table confirms that dual-oscillator noise systematically reduces $t^*_{LL}$ compared to both white noise ($t^*_W$) and single-oscillator colored noise ($t^*_L$) in both near the boundary of the parameter domain explored in this work parametric regime ($\Lambda=0.48$) and more stable regime ($\Lambda=0.451$). 
\begin{table}[h]
	\centering
	\caption{ Parameter regimes and Mpemba crossing times $t^*$. Weak noise: $\gamma_L=0.09\omega_a$; moderate: $\gamma_L=0.05\omega_a$. $t_{LL}^*$ correspond to crossing time of colored noise applied on both oscillators. All times in units of $\omega_a^{-1}$. Meanwhile, $No$ means no crossing was observed with these parameters}
	\begin{tabular}{lcccc}
		\hline
		Regime & $\Lambda$ &  $t^*_\mathrm{W}$ & $t_L^*$  & $t_{LL}^*$\\
		\hline
		Weak Colored Noise & 0 & $No$ & $No$ & $No$ \\
		Weak Colored Noise & 0.451 & $No$ & 1.651 & 1.407 \\
		Moderate Colored Noise & 0.451 & $No$ & 1.508 & 1.201 \\
		Weak Colored Noise & 0.48 & 0.905 & 0.895 & 0.865 \\
		Moderate Colored Noise & 0.48 & 0.905 & 0.879 & 0.799 \\
		\hline
	\end{tabular}
	\label{tabletimes}
\end{table}
	
\begin{figure}
	\begin{center}
		\includegraphics[width=7.5cm,height=5cm]{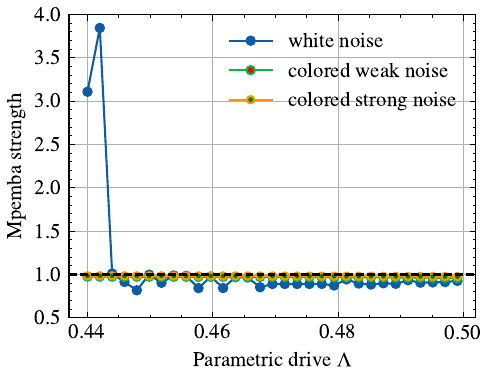}
	\end{center}
	\caption{(color online) Mpemba strength, defined as $M = |c^{(h)}_1|/|c^{(c)}_1|$, as a function of the parametric drive $\Lambda$ for the system under Gaussian white noise (blue circles), double weak Lorentzian colored noise (green circles), and double moderate Lorentzian colored noise (orange circles). For most of the stable parameter range, $M$ remains close to unity, indicating that the Mpemba effect is not very strong. However, colored noise extends the Mpemba regime to smaller values of $\Lambda$, where the white-noise case does not show a crossing.}\label{fig5}
\end{figure}

We define the Mpemba strength as $M = |c_1^{(h)}|/|c_1^{(c)}|$, so that $M<1$ indicates a Mpemba effect. Figure. (\ref{fig5}) shows the variation of Mpemba strength versus $\Lambda$ for this system under white, weak colored and moderate strong noise. As can be seen, the Mpemba strength remains close to unity over most of the stable parameter range, indicating that colored noise only weakly modifies the slow-mode overlap. Sharp spikes larger than $1$ for some certain values far from value of $\Lambda = 0.5$ under Gaussian white noise are adapted to previous obtained results of crossing time. These results confirm that the parametric drive under white noise reflects strong sensitivity of the spectral projection in this regions. The eigenvector structure changes enough to produce large amplification of the Mpemba criterion for these parametric drive values. 

The enhancement of the Mpemba effect by Lorentzian colored noise can be understood from the modification of the relaxation pathways in covariance space. In the white-noise case, relaxation is governed by the deterministic drift matrix together with a Markovian diffusion process. Introducing colored noise adds finite environmental memory through auxiliary Ornstein--Uhlenbeck variables, leading to an enlarged dynamical system.
The resulting memory effects do not qualitatively alter the relaxation mechanism responsible for the Mpemba effect. Instead, they modify the spectral decomposition of the relaxation dynamics and redistribute the projections of the initial states onto the relaxation modes. In particular, the overlap of the hot state with the slowest relaxation manifold is reduced, allowing the distance to the steady state to decay more rapidly at intermediate times.
This explains why colored noise generally decreases the crossing time while preserving the same qualitative relaxation scenario observed in the white-noise case.

When colored noise is applied to both oscillators, memory-assisted relaxation channels are introduced simultaneously in the two coupled subsystems. The resulting extended drift matrix contains additional couplings between the physical quadratures and the auxiliary Ornstein--Uhlenbeck variables associated with each oscillator. Consequently, the redistribution of spectral weights occurs in a larger portion of the phase space, leading to a stronger suppression of the slow-mode contribution and therefore to shorter crossing times than in the single-channel colored-noise configuration.
	 
\begin{figure}
		\begin{center}
			\includegraphics[width=8.5cm,height=4cm]{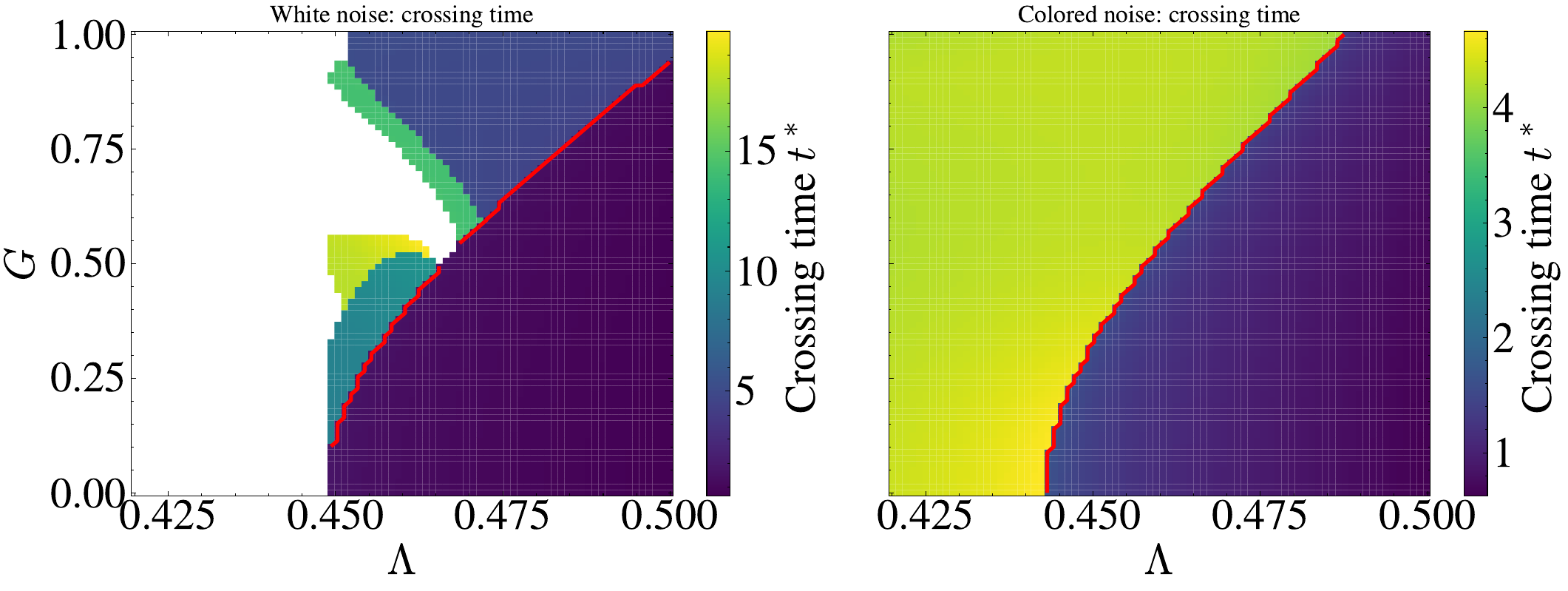}
		\end{center}
		\caption{ (color online) Crossing time $t^*$ in the $(\Lambda,G)$ plane for white noise (left) and Lorentzian colored noise (right). The color scale shows the first crossing time between the hot and cold relaxation curves. The white region for the white noise shows no crossing time occurs up to $t_{max}=20$. Red curve is the contour $t^*=2$ separates delayed crossing time region from interested Mpemba region $t^*<2$.}\label{fig6}
	\end{figure}

Figure~\ref{fig6} presents the crossing time $t^*$ of the distance curves between the hot and cold initial states as a function of the parametric amplification strength  $\Lambda$ and the coupling strength $G$, for both white and one moderately strong Lorentzian colored noise. The left panel corresponds to white noise, while the right panel shows the colored noise case. In each panel, the color scale represents the first time at which the distance of the hot state becomes smaller than that of the cold state, i.e., the Mpemba crossing time $t^*$. 

The white region in the white-noise panel indicates that no	crossing is observed within the chosen observation window.	Comparison with the colored-noise case demonstrates that environmental memory extends the parameter region in which the Mpemba effect can be observed. Red curve is the contour $t^*=2$ separates delayed crossing time region from interested (fast) Mpemba region $t^*<2$.
As can be seen,  $\Lambda$ approaches the boundary of the stable parameter regime considered here, $0.5$, the relaxation becomes increasingly sensitive to parameter variations. The dependence on $G$ shows that stronger coupling can either promote or suppress the crossing time, depending on the noise type and on how the coupling redistributes relaxation pathways between the modes. The dependence on $\Lambda$ again indicates the fact that the parametric drive modifies the effective normal-mode frequencies and the relaxation rates of the coupled oscillator system.

In this representation, one can identify parameter regions where the Mpemba effect is robust, as opposed to regions where the crossing is delayed beyond the chosen observation window. This type of phase diagram is useful for guiding experiments, since it identifies the parameter combinations that maximize the likelihood of observing accelerated relaxation from the hotter initial state. 

\section{Conclusion} \label{sec:conclusion}
We have investigated the Mpemba effect in a system of two linearly coupled harmonic oscillators, one of which is parametrically driven and dissipatively coupled to a thermal environment. Within the covariance matrix formalism, we analyzed the relaxation dynamics under Gaussian white noise and Lorentzian noise, and used both the Frobenius distance and the slow-mode projection to identify Mpemba crossings.
Our results show that the parametric drive is the main factor controlling anomalous relaxation. As the parametric drive approaches the edge of the stable parameter regime considered here, the Mpemba crossing time decreases due to changes in the spectral projections and relaxation-mode structure. Colored noise provides an additional enhancement, especially when it acts on both oscillators, but its effect remains secondary compared with the deterministic spectral structure of the drift matrix. We also found that colored noise extends the parameter region where the Mpemba effect persists, particularly for smaller values of the parametric drive where white noise alone may not produce a crossing.
	
The initial squeezing used to prepare the hot state helps place the two initial states in a suitable nonequilibrium regime, but it does not qualitatively alter the relaxation mechanism. Overall, our analysis indicates that parametrically driven coupled oscillators offer a controllable and experimentally accessible platform for studying and engineering anomalous thermal relaxation. These results may be relevant to cavity optomechanics, superconducting parametric amplifiers, and driven nanomechanical systems, where both parametric control and environmental memory can be used to tailor relaxation pathways. In such platforms, finite-correlation-time noise can be generated using auxiliary damped resonators, filtered noise injection, or feedback-based reservoir engineering, providing possible routes for experimental observation of the predicted colored-noise enhancement of the Mpemba effect.

\appendix
\section{Routh-Hurwitz criterion}
\label{App:Routh_Hurwitz}
\begin{figure}
	\begin{center}
		\includegraphics[height=5cm, width=7.2cm]{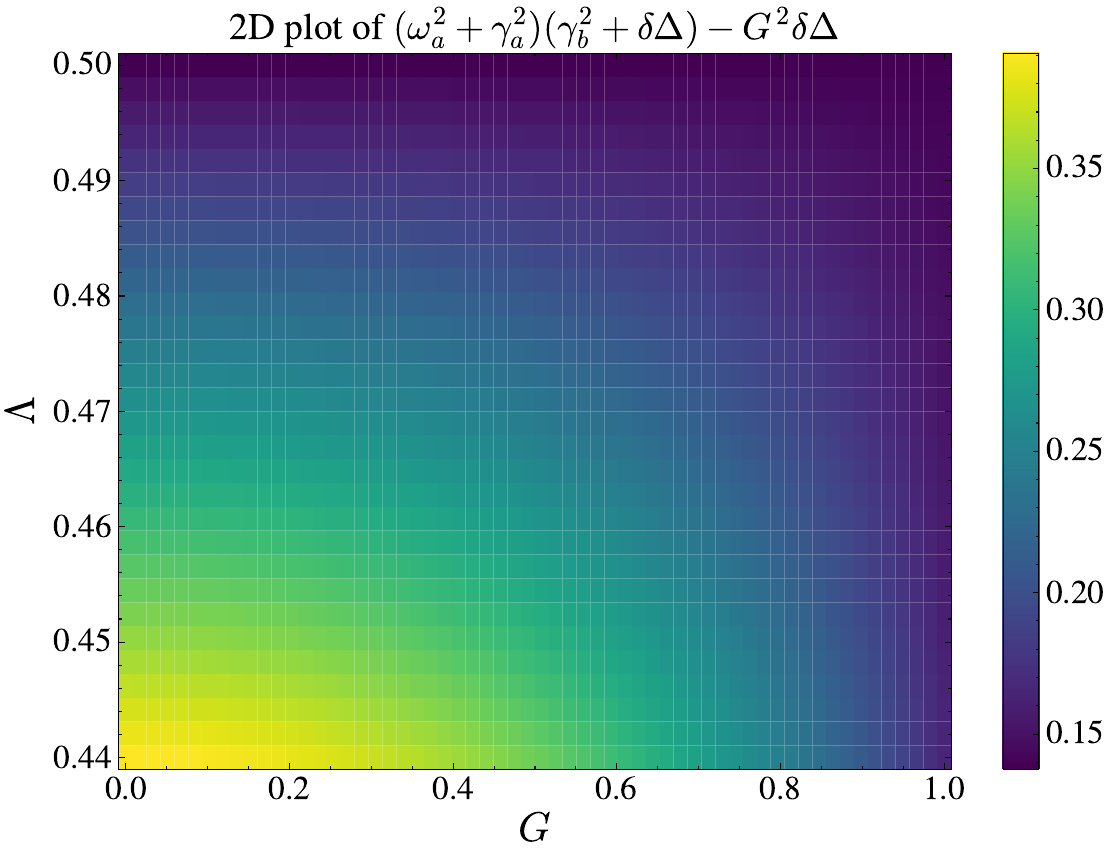}
	\end{center}
	\caption{(color online) The stability condition $(\omega_a^2 + \gamma_a^2)(\gamma_b^2 + \delta \Delta) - G^2 \delta \omega_a  $ vs $\Lambda$ and $G$ while $\omega_a = \Delta_b =1$, $\gamma_a = \gamma_b = 0.45 \omega_a$.}\label{fig7}
\end{figure}

For the sake of clarity, the stability condition $(\omega_a^2 + \gamma_a^2)(\gamma_b^2 + \delta \Delta) - G^2 \delta \omega_a >0$ obtained from Routh-Hurwitz criterion has been checked and shown in Fig. \ref{fig7} for different values of $\Lambda$ and $G$. These results confirm the stability of the system over the ranges of $\Lambda$ and $G$ that were used in our simulations.
Meanwhile, for the parameters used throughout this work, $ \max_i \mathrm{Re}(\lambda_i) = -0.45$, confirming that all simulated parameter points remain within the stable regime. 

\section{Squeezing parameters effect}
\label{App:Squeezing parameters}
\begin{figure}[H]
	\begin{center}
  		\includegraphics[width=8.5cm,height=3.6cm]{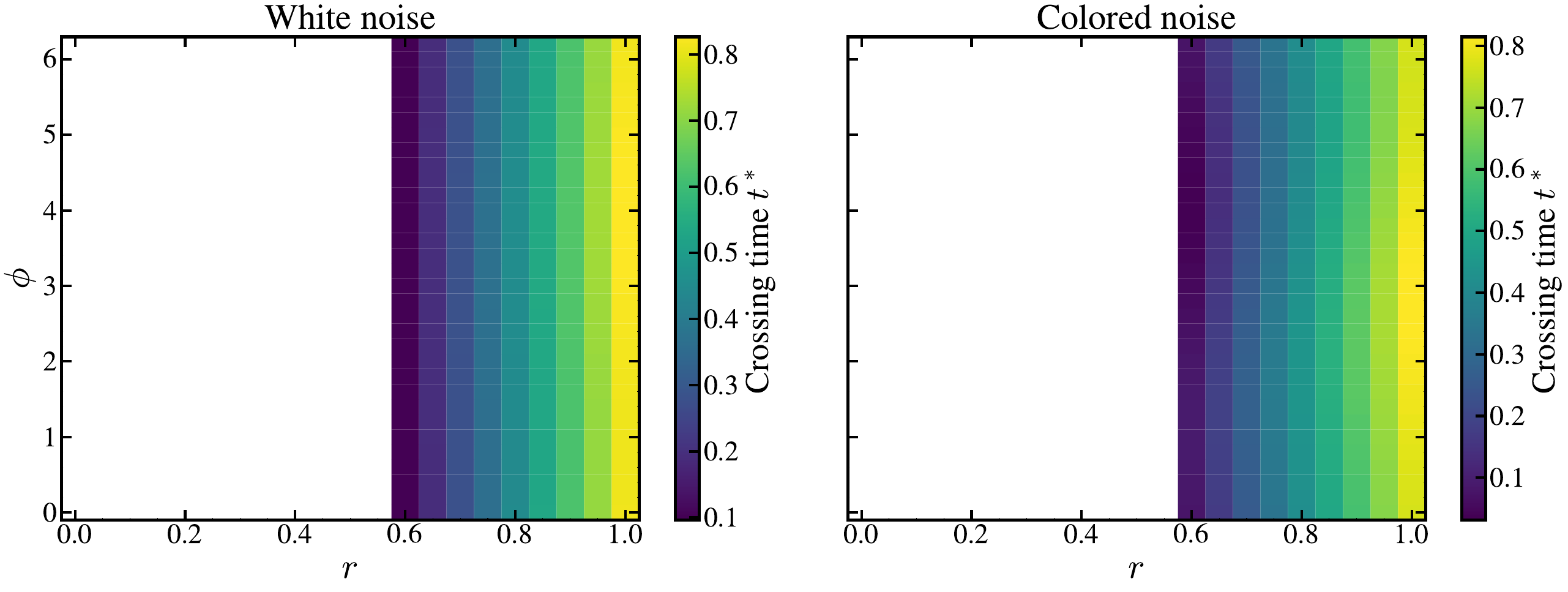}
  	\end{center}
  	\caption{ Mpemba crossing time $t^\ast$ as a function of the squeezing	strength $r$ and squeezing angle $\phi$ for Gaussian white noise and weak Lorentzian colored noise. The parameters are $\omega_a=\Delta_b=1$, $\gamma_a=\gamma_b=0.45\omega_a$, $\Lambda=0.48$, and $G=0.2\omega_a$. No crossing is observed within the numerical observation window for $r\lesssim0.6$ for these parameters; this apparent threshold is a state-preparation crossover, not a dynamical stability threshold.}\label{fig8}
\end{figure}

Figure \ref{fig8} shows the Mpemba crossing time as a function of the squeezing parameters $r$ and $\phi$ for both white noise and weak Lorentzian colored noise. As can be seen, to see the crossing time we should apply the parameter $r \ge 0.6$. 
To clarify the role of the squeezing strength, consider the $2\times2$ covariance block of oscillator $a$ in the
initial thermal state,
\begin{equation}
	\sigma_{{\rm th},a}^{(i)} = v_a^{(i)}I_2,
	\qquad
	v_a^{(i)} = \bar n_a^{(i)}+\frac{1}{2},
\end{equation}	
where $i=c,h$. Here $v_a^{(i)}$ is the common variance of the two thermal quadratures, $\langle(\Delta x_a)^2\rangle= 	\langle(\Delta p_a)^2\rangle=v_a^{(i)}$. Equivalently, $v_a^{(i)}$ is the doubly degenerate ordinary eigenvalue of the isotropic thermal covariance block.
The same local squeezing operation is applied to oscillator $a$ in both the hot and cold initial preparations, whereas oscillator $b$ remains initially thermal. The corresponding squeezed covariance block of oscillator $a$ has the principal variances 
\begin{equation}
	\nu_{a,\pm}^{(i)}=v_a^{(i)}e^{\pm2r},
\end{equation}
which are the ordinary eigenvalues of the squeezed covariance block. The resulting covariance anisotropy is therefore
\begin{equation}
	\nu_{a,+}^{(i)}-\nu_{a,-}^{(i)}	= 2v_a^{(i)}\sinh(2r).
\end{equation}
Thus, increasing $r$ enhances the eccentricity of the initial covariance ellipse of oscillator $a$ and changes the direction of the initial covariance deviation in the vectorized covariance space. Through the intermode coupling, this initially local deformation influences the subsequent two-mode relaxation dynamics. Consequently, it modifies the projections of the hot and cold initial states onto the relaxation modes. In particular, a sufficiently large squeezing-induced deformation can reduce the hot-state overlap with the slow relaxation manifold relative to that of the cold state, thereby favoring the spectral Mpemba condition $|c_1^{(h)}| < |c_1^{(c)}|$.
The threshold-like behavior in $r$ is therefore neither a dynamical instability nor a phase transition. Rather, it is a
state-preparation crossover determined by the relative orientation of the initial covariance deviations and the slow relaxation manifold.

The squeezing angle $\phi$ has a different role: it rotates the squeezing ellipse of oscillator $a$ in its local $(x_a,p_a)$ quadrature plane while leaving its principal variances unchanged. Consequently, varying $\phi$ mainly redistributes the initial fluctuations among $\sigma_{x_ax_a}$, $\sigma_{p_ap_a}$, and $\sigma_{x_ap_a}$, whereas $r$ changes the magnitude of the anisotropy itself. For the parameter regime considered here, the relevant slow relaxation modes are only weakly sensitive to this local rotation, which accounts for the weak dependence of the crossing time on $\phi$. The residual angular dependence arises because the parametric drive, intermode coupling, and unequal bath occupations break exact rotational symmetry in the full four-dimensional phase space.
The crossover value $r\simeq 0.6$ is not universal. It depends on the drift parameters, the bath occupations, the noise-correlation parameters, and the distance used to identify the crossing. Changing $\Lambda$, $G$, $\gamma_{a,b}$, $\bar n_{a,b}$, or the colored-noise
parameters changes the relaxation eigenvectors and the stationary covariance, and therefore shifts the minimum squeezing strength required to observe a crossing.

\section*{Declarations} 
\subsection*{Funding} No funding was received for this research. 
\subsection*{Availability of data and materials} Data sharing not applicable to this article as no datasets were generated or analysed during the current study.
\subsection*{Competing interests} The authors declare no competing interests.
\subsection*{Authors' contributions} A.P. and M.R. contributed equally to the conception and development of
the work, numerical analysis, interpretation of the results, and preparation of the manuscript. M.A.-R. contributed to the development of the work, analysis and interpretation of the results, and revision of the manuscript. All authors read and approved the final manuscript.

\end{document}